\documentclass[preprint2,longabstract]{aastex}

\shorttitle{G313.3+00.3: A New Planetary Nebula}
\shortauthors{Cohen et al.}

\begin{document}
\title{G313.3+00.3: A New Planetary Nebula discovered by 
the Australia Telescope Compact Array and the {\it Spitzer} Space Telescope}

\author{Martin Cohen\altaffilmark{1},Anne J. Green\altaffilmark{2},Mallory S.E. Roberts\altaffilmark{3},
Marilyn R. Meade\altaffilmark{4},Brian Babler\altaffilmark{4},\\
R\'emy Indebetouw\altaffilmark{5},Barbara A. Whitney\altaffilmark{6},Christer Watson\altaffilmark{7},Mark Wolfire\altaffilmark{8},
Mike J. Wolff\altaffilmark{6},\\
John S. Mathis\altaffilmark{4},and Edward B. Churchwell\altaffilmark{4}}

\altaffiltext{1}{Radio Astronomy Laboratory, University of California,\\ 
Berkeley, CA 94720; {\bf mcohen@astro.berkeley.edu}}
\altaffiltext{2}{School of Physics, University of Sydney, NSW 2006, Australia}
\altaffiltext{3}{Dept. of Physics, McGill University, Montreal, Canada}
\altaffiltext{4}{Dept. of Astronomy, University of Wisconsin, Madison,\\ 
WI 53706}
\altaffiltext{5}{Astronomy Dept. University of Virginia, Charlottesville,\\
VA 22904}
\altaffiltext{6}{Space Science Institute, Boulder, CO 80303}
\altaffiltext{7}{Dept. of Physics, Manchester College, North Manchester,\\ 
IN 46962}
\altaffiltext{8}{Dept. of Astronomy, University of Maryland, College Park,\\ 
MD 20742}

\begin{abstract}
We present a new planetary nebula, first identified in images from the Australia 
Telescope Compact Array, although not recognized at that time.  Recent observations
with the {\it Spitzer} Space Telescope during the GLIMPSE Legacy program have
rediscovered the object.  The high-resolution radio and infrared images enable the
identification of the central star or its wind, the recognition of the radio 
emission as thermal,
and the probable presence of polycylic aromatic hydrocarbons in and around the source.
These lead to the conclusion that G313.3+00.3 is a planetary nebula.  This object is 
of particular interest because it was discovered solely through radio and mid-infrared 
imaging, without any optical (or near-infrared) confirmation, and acts as a proof of
concept for the discovery of many more highly extinguished planetary nebulae.
G313.3+00.3 is well-resolved by both the instruments with which it was identified,
and suffers extreme reddening due to its location in the Scutum-Crux spiral arm.
\end{abstract}


\section{Introduction}
The GLIMPSE (Galactic Legacy Infrared Mid-Plane Survey Extraordinaire) 
Legacy program (Churchwell et al. 2004) is surveying 220~deg$^2$ of the plane
in all four bands of the Infrared Array Camera (IRAC: Fazio et al. 2004) aboard the 
{\it Spitzer} Space Telescope (Werner et al. 2004).  GLIMPSE is generating a wealth of 
detailed imagery of the plane at mid-infrared (MIR) wavelengths.  The greatest
scientific yield from any such survey occurs when multiwavelength data are compared. 
Detailed panoramic comparisons of areas of the Galactic plane in the radio continuum
and MIR were undertaken by Cohen \& Green (2001), who examined
8~deg$^2$ in the {\it l}~=~312$^\circ$ region using matched resolution imagery.
These authors investigated the detailed 843-MHz inventory of this field made by
Whiteoak et al. (1994) with the Molonglo Observatory Synthesis Telescope (MOST;
Mills 1981; Robertson 1991).  The Midcourse Space eXperiment (MSX: Price et al. 2001)
detected a substantial number of H{\sc ii} regions, of diverse morphology, in its MIR 
survey of the Galactic Plane, which Cohen \& Green (2001) used to calibrate the ratio 
of 8.3-$\mu$m to 843-MHz flux densities, $F_{8.3 \mu m}/S_{843 MHz}$, as a discriminant 
between thermal and nonthermal radio sources.  We are engaged in a detailed comparison 
of the GLIMPSE images of the same {\it l}~=~312$^\circ$ field previously studied with
MSX and the MOST.  

This paper presents high-resolution radio continuum and {\it Spitzer} images of G313.3+00.3, 
an object which appears as a nebulous ring with a central star.  The radio images,
together with a new 843-MHz image, have been used to estimate a spectral index for the
source, which implies that the object is thermal (\S2).  Previous infrared (IR) and optical 
information available in the literature constitute \S3.  
IR images of the nebula taken with the IRAC are discussed in \S4, and alternative explanations 
for the nebula are considered in \S5, after which (\S6) we conclude that the source 
is an extremely highly reddened planetary nebula (PN).

\section{Radio continuum observations}
The original impetus for carrying out high-resolution radio continuum
images of this region was to search  for posssible counterparts to an
unidentified Galactic plane $\gamma$-ray source. The initial radio
survey used to pinpoint the target region was the first epoch Molonglo
Galactic Plane Survey (MGPS1; Green et al. 1999) from the MOST.
That 843-MHz continuum observation has a resolution of 
$43^{\prime\prime}~\times~43^{\prime\prime}~$cosec$\delta$ and a 1$\sigma$ sensitivity 
of $1-2$ mJy beam$^{-1}$. The MOST image used
in the present paper to calculate flux densities is a mosaic (J1430M60) of the
second epoch survey MGPS2 (Green 2002) which is available at the
website http://www.astrop.physics.usyd.edu.au/SUMSS. The observational
parameters are essentially unchanged from MGPS1.

The targeted radio continuum images were part of the project
to search for the radio counterpart of the EGRET source GeV $J1417-6100$ 
(Roberts et al. 1999). Observations were made at 13 and 20~cm
with the Australia Telescope Compact Array (ATCA; Frater et al.1992) 
in three configurations: 1.5A (1998 January 11), 0.75A
(1998 April 22), and 0.375 (1998 March 28). In addition, high-resolution 20-cm 
data were included from three observations taken in 1999 and 2001 in the 6D, 
0.375, and 0.75C configurations by Froney Crawford and Simon Johnston (these
are detailed by Roberts et al. (2001)). The ATCA can observe two
simultaneous frequency bands and 1384 and 2496~MHz were selected. For
the 1.5A array, the lower frequency was slightly different, at
1344~MHz. For all observations a bandwidth of 128~MHz was used and
full Stokes parameters were recorded. For most of the analysis in
Roberts et al. (1999), images were used which did not include the
baselines to the 6-km antenna.

High-resolution images including the 6-km antenna of the ATCA were
also made, using the MIRIAD (Sault et al. 1996),
maximum entropy method to minimize artefacts. The ring source
presented in this paper is shown in the high-resolution 13-cm image of
Figure~3 of Roberts et al. (1999), but there is no discussion on its
nature. This image has an angular resolution of $2.8^{\prime\prime}~
\times~2.5^{\prime\prime}$ and an rms noise of 0.1~mJy~beam$^{-1}$. The
high-resolution 20-cm image presented in this paper has an angular
resolution of $5.1^{\prime\prime}~\times~4.6^{\prime\prime}$ and an rms
noise of 0.1~mJy~beam$^{-1}$. In both the high-resolution images a
shell is clearly seen. In the low resolution images, the source is
seen as slightly extended which allows an accurate measure of the
total flux density for spectral index calculations.  These total flux
densities at the three frequencies appear in Table~\ref{flux}.  
A formal regression yields a slope of $0.000\pm0.002$, and an essentially 
constant flux density of $25\pm5$~mJy, with $\chi^2$~=~4.3, consistent
with a flat spectrum from thermal bremsstrahlung.  Note that, at these frequencies, 
the Galaxy is optically thin so the issue of synchrotron self-absorption 
along the line-of-sight does not arise.

\begin{figure}
\plotone{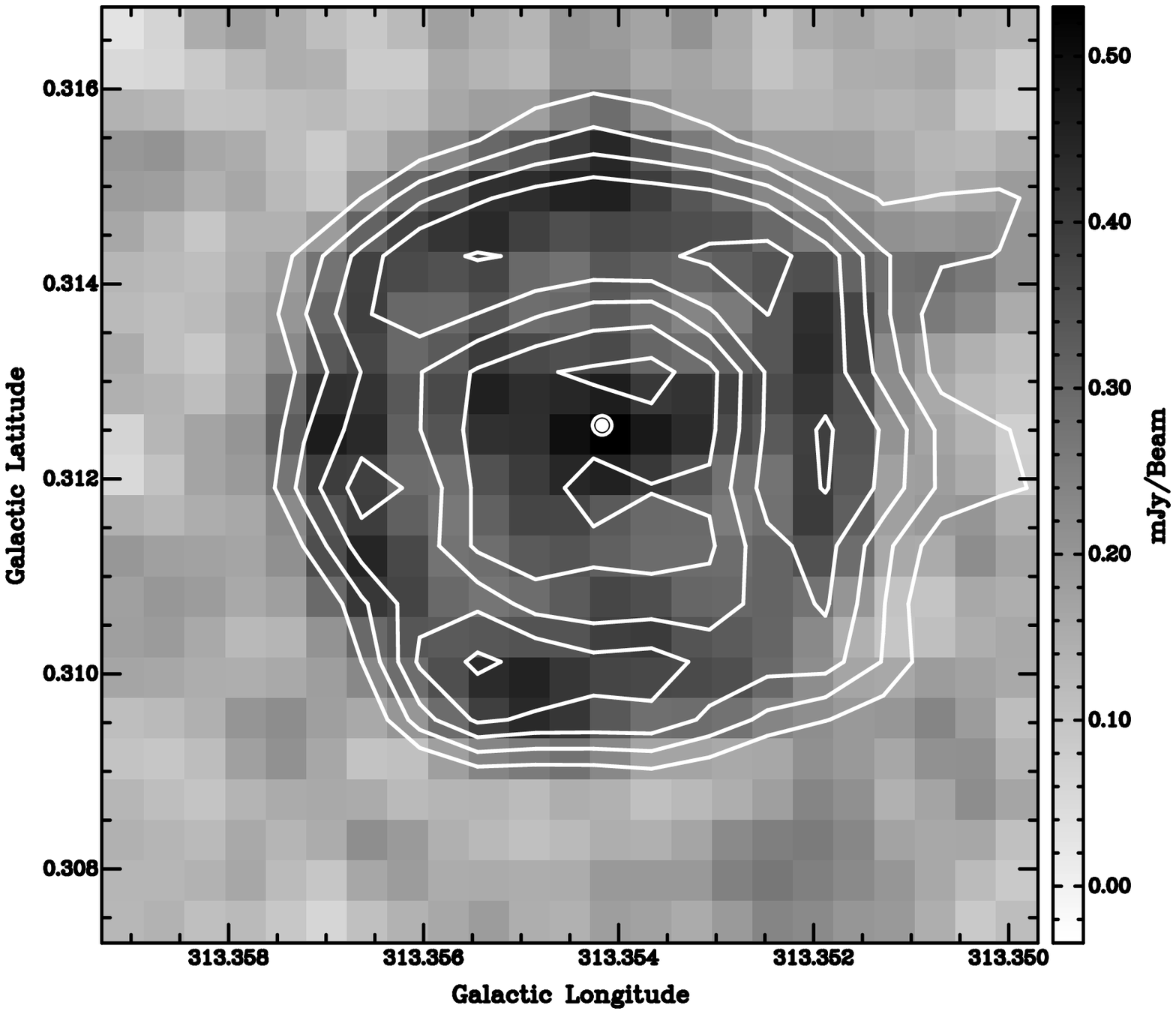} 
\figcaption{High-resolution 13-cm greyscale image of G313.3+00.3 overlaid by 20-cm contours
at 0.6, 0.8, 1.0, 1.2, 1.5~mJy~beam$^{-1}$.
The position of the central star coincides with the white dot
surrounded by a small white circle. \label{13_20}}
\end{figure}

Figure~\ref{13_20} presents the high-resolution 13-cm image of the nebula
overlaid by contours of high-resolution 20-cm emission.  A central point source is 
detected at both frequencies although its peak is better determined at 13~cm:
$14^h18^m27.489^s\pm0.007^s$, $-60^{\circ}47^{\prime}10.49^{\prime\prime}\pm0.06^{\prime\prime}$ (J2000),
corresponding to {\it l=}313.3541$^\circ$, {\it b=}+0.3125$^\circ$.  It lies at the 
geometric center of the nebula.  We identify this as the location of the
central star of the nebula.  

\begin{figure}
\plotone{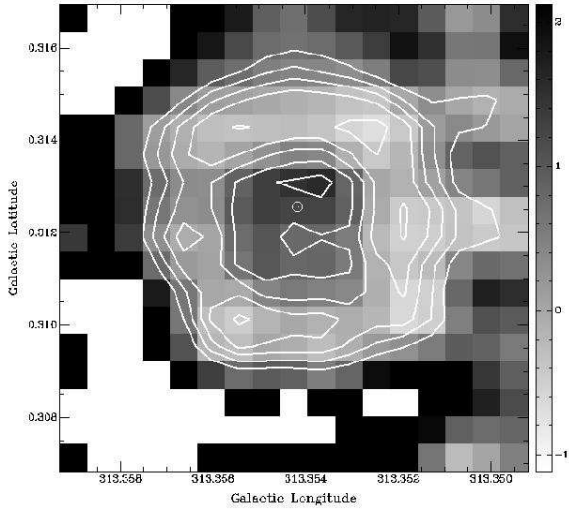} 
\figcaption{Radio spectral index map of the nebula between 13 and 20 cm, in the range
$\alpha$ from about $-1$ to +2. The star's position is within the small white circle.
20-cm contours are as in Fig.~\ref{13_20}. White pixels represent indeterminate
spectral indices.\label{spind}}
\end{figure}

Figure~\ref{spind} shows the spatial variation of spectral index between 13 and
20~cm across the nebula and at the stellar location.  This map was made
by matching the 13-cm image to the resolution and pixel grid of the
20-cm map.  The ratio of these two images was scaled to 
represent the spectral index, $\alpha$, where S$_\nu$~$\propto$~$\nu^{\alpha}$.
At the star, $\alpha\approx1.6$, which is somewhat flatter than a Rayleigh-Jeans slope
for emission from optically thick gas. 
At the location of the central star is a 13-cm point source with a flux density 
of 1~mJy.  It is difficult to extract a 20-cm counterpart because locally there is
extended emission of comparable brightness to the stellar peak.  The formal
fit yields 2.6~mJy although this peak is $\sim$1.5 beams away from the 
well-determined 13-cm peak.  If this does represent the star at 20~cm then
the spectral index is $\sim$1.7, with probable uncertainty large enough to
include an index of 2, i.e. emission in the Rayleigh-Jeans regime of a hot
object.  
The region immediately surrounding the star is occupied by nebular gas with an 
average index of $0.75\pm0.08$.  The outer annulus of the nebula has  
$\alpha$~=~$-0.09\pm0.06$, indicative of thermal (free-free) emission.

\section{Infrared and optical data from the literature}
The first detection of a MIR source at this location was made by {\it IRAS}.
The {\it IRAS} Point Source Catalog lists {\it IRAS}~$14147-6033$, whose original (B1950)
position corresponds to $14^h18^m27.4^s\pm2.6^s$, 
$-60^{\circ}47^{\prime}08^{\prime\prime}\pm16^{\prime\prime}$ (J2000).  It was 
detected only at 25~$\mu$m with a flux density below 1~Jy.  Consequently, the
object has never been classified, because neither color nor color-color
criteria could be applied.

During their MIR and radio study of the {\it l}~=~312$^\circ$ region
Cohen \& Green (2001) used MSX imagery of the Galactic plane with
20$^{\prime\prime}$ resolution from the ``CB-02" data collection
events.  The nebula is faint but was detected in these images, although it
was essentially unresolved.  Additional, deeper MSX data are now available 
(the ``CB-03" data) which are roughly 5$\times$
deeper than the CB-02 images.  The nebula is not listed in the MSX Point Source Catalog 
(Ver. 2.3: Egan et al. 2003), although it was covered in the CB-03 field centered near 
{\it l}~=~315$^\circ$ (Price et al. 2001). In this deeper 8.3-$\mu$m image it 
appears as a resolved object about $30^{\prime\prime}\times30^{\prime\prime}$, 
i.e. with a deconvolved size of $\approx22^{\prime\prime}$.
The same source is also detected in the 14.6-$\mu$m CB-03 image of this same
field.  These MIR flux densities are collected into Table~\ref{flux}.

A search of the 2MASS archives and images shows no near-infrared (NIR) counterpart
to the nebula.  Likewise, no southern broadband optical ($B_j, R, I$) image, 
from the UK Schmidt Telescope/SuperCosmos surveys shows a counterpart to
the nebula.  However, there may be an extremely faint counterpart 
in the highly sensitive narrowband Southern H$\alpha$ Survey of the Galactic 
plane (SHS: Parker \& Phillipps 1998,2003), which can serve at least as an upper
limit to H$\alpha$ brightness (see \S5).  These attributes suggest a highly reddened 
and/or distant object.

\section{New {\it Spitzer} infrared images}
Cohen \& Green (2001) established a distinction between thermal and nonthermal radio emission 
by comparing MIR and radio counterparts of many different types of H{\sc ii} region
in the {\it l}~=~312$^\circ$ area.  GLIMPSE offers a deeper exploration of this field 
and could potentially unravel the most complex areas in which MIR-emitting thermal
filaments are juxtaposed with regions known to be nonthermal from their 
radio emission.  When we examined the IRAC Band-4 (8.0-$\mu$m) mosaic images
in the vicinity of the EGRET source $J1417-6100$ (Roberts et al. 1999) we immediately 
saw an IR ring nebula and a central star, where the lower-resolution MSX 
data showed just a faint blob at 8.3~$\mu$m.  

\begin{figure}
\plotone{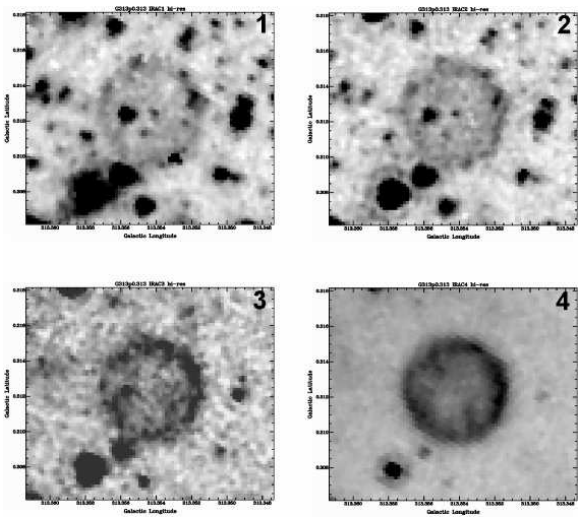} 
\figcaption{Montage of IRAC high-resolution images of the ring in the 4 bands.
Pixel size is 0.6$^{\prime\prime}$.  Diffraction patterns can be seen around
the brighter point sources.\label{irac1-4}}
\end{figure}

The individual IRAC frames were calibrated by the {\it Spitzer} Science Center
(SSC), and further processed by the GLIMPSE pipeline to remove instrumental 
artifacts such as column pulldown and banding (Hora et al. 2004a).  Point sources 
were extracted from each frame using a modified version of DAOPHOT (Stetson 1987), 
and cross-referenced using the SSC bandmerger. Mosaics were created using the Montage 
package\footnote{available at http://montage.ipac.caltech.edu}.  The final GLIMPSE 
image format has 0.6$^{\prime\prime}$ pixels and appears in Figure~\ref{irac1-4}, 
which illustrates the four IRAC images of this nebula, which becomes progressively 
more conspicuous with increasing wavelength.  The mosaic images conserve surface 
brightness.  Spatially integrated, background-subtracted fluxes for 
the whole nebula were extracted at each wavelength (see Table~\ref{flux}) after 
correcting for obvious point sources within the ring or lying near its rim.  The flux
calibration of GLIMPSE is cumulatively checked after completion of each segment
of the survey against a network of calibrators established in the survey area 
along the Galactic Plane (Cohen 2004).  At present, the observed absolute flux 
densities of a sample of about 70 early A-dwarf and 60 K-giant calibrators show 
agreement with predictions to within $\sim$3\% in all IRAC bands (Cohen 2004).

At the center of the ring is a local
peak in the IRAC Band~1 (3.6~$\mu$m) and Band~2 (4.5~$\mu$m) images, which is
swamped by bright nebulosity in Band~3 (5.8~$\mu$m) and Band~4 (8.0~$\mu$m).
This peak pixel is at {\it l=}313.3541$^\circ$, {\it b=}+0.3125$^\circ$
and is clearly the IR counterpart of the central point source detected in the radio.
Its designation in the GLIMPSE archive is: SSTGLMA G313.3541+00.3124.
Two circles with radii of 12.6$^{\prime\prime}$ and 7.6$^{\prime\prime}$, 
centered on this NIR peak, provide good fits to the outer and inner edges of 
the ring, validating the choice of the central source as the exciting star.  
PSF-fitting photometry of the central star was extracted from the 3.6 and 4.5-$\mu$m 
images, above the locally determined nebular emission inside the ring. There is no 
noticeable central star in the 5.8 or 8.0-$\mu$m images.

\begin{figure}
\plotone{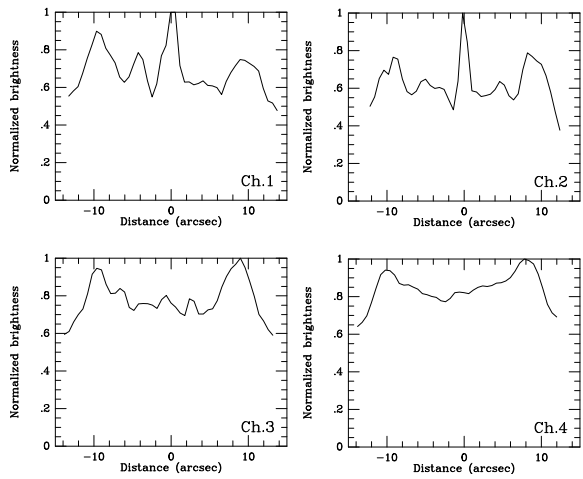} 
\figcaption{Brightness profiles through the central star's position in the four
IRAC high-resolution images.  Negative distances from the center are to the
south of the star.\label{prof1-4}}
\end{figure}

Profiles in Galactic latitude through the center of the ring are presented in
Figure~\ref{prof1-4}.  Each normalized profile was constructed from the average 
of the three adjacent spatial slices centered on, and flanking, the position
of the peak Band1 pixel of the central star.  Negative distances from the
center correspond to southward latitudes from the star. 

\begin{figure}
\plotone{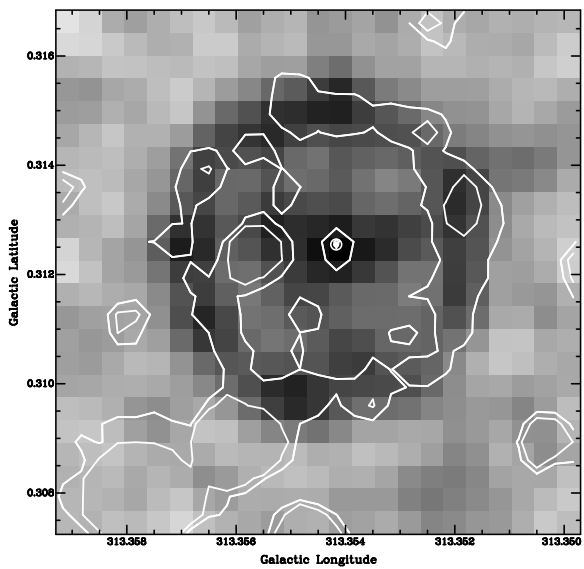} 
\figcaption{13-cm high-resolution greyscale image overlaid by white contours of
4.5-$\mu$m emission at values of 3, 4, 5~MJy~sr$^{-1}$.  The lowest contour 
is 4$\sigma$ above the background.
The star is seen at both wavelengths in the center of the clumpy shell and is
shown as in Fig.~\ref{13_20}. Note the good match in spatial extent of the
outer boundaries in the IR and radio maps.\label{13cm_i2}}
\end{figure}

\begin{figure}
\plotone{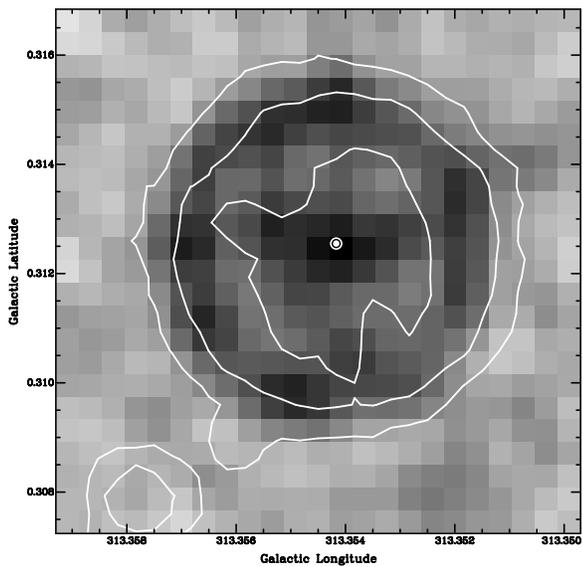} 
\figcaption{13-cm high-resolution greyscale image overlaid by white contours of
8.0-$\mu$m emission at values of 34, 42.5~MJy~sr$^{-1}$.  The lowest contour 
is 4$\sigma$ above the background,
and corresponds to 34~MJy~sr$^{-1}$, as does the third contour.  The second
contour represents the bright ridgeline of the IR ring (42.5~MJy~sr$^{-1}$).
Star shown as in Fig.~\ref{13_20}. Note the greater spatial extent of the IR
outer boundary compared with the radio emission.\label{13cm_i4}}
\end{figure}

Figures~\ref{13cm_i2} and \ref{13cm_i4} compare the high-resolution 13-cm with the
4.5-$\mu$m and 8.0-$\mu$m images. In each IR band the lowest contour corresponds to
4$\sigma$ above the local background level. Only the 4.5-$\mu$m IRAC band should contain
no significant contribution from PAH emission. Radiation in this band might
represent line emission from ions (e.g. [Mg{\sc iv}]), or molecular gas
(e.g. ro-vibrational H$_2$ lines, 0-0~S(8)-S(11)), thermal emission from free-free 
continuum, or from heated dust grains.

The good match in nebular extent and morphology between the 4.5-$\mu$m and radio 
maps would suggest that both these images trace ionized gas.  By contrast, the 
8.0-$\mu$m emission clearly extends beyond the radio thermal emission, consistent 
with the presence of polycyclic aromatic hydrocarbon (PAH) emission in the 
8.0-$\mu$m bandpass.  This would be expected if the PAHs resided in the 
photodissociation region (PDR) surrounding the ionized gas (e.g. Cohen \& Green 2001).
This 8.0-$\mu$m extension beyond the ionized gas could also be consistent with the
presence of molecular hydrogen emission lines in IRAC Band~4.

\section{The nature of the nebula}
We will calculate the mass and distance of the nebula, after determining
the contribution of recombination lines, and free-free and bound-free continua,
to the IRAC fluxes, based on the radio free-free fluxes, and estimating the
interstellar extinction from an upper limit to the H$\alpha$ flux, and from
extinction maps based on DIRBE data.

We have calculated the contributions suggested by the radio emission 
at 13~cm, assumed to be thermal, to the H-line emission within each 
of the IRAC filters.  The strongest expected recombination lines are
Pf$\gamma$ and Pf$\delta$ in IRAC1; Br$\alpha$ and Pf$\beta$ in IRAC2;
Hu$\gamma$ in IRAC3; and Pf$\alpha$ and Hu$\beta$ in IRAC4.
All the 13-cm emission was attributed to free-free, using emissivities 
from Storey \& Hummer (1995) for T$_e$=10$^4$~K (the density is almost 
unimportant), approximating the radio free-free Gaunt factor by 
Osterbrock (1989: eqn. 4.30), and ignoring the effects of any dust
mixed with the ionized gas.  
The free-free continuum within the IRAC bands was computed with IR 
Gaunt factors from Beckert et al. (2000), and the bound-free 
component was estimated following Seaton (1960) for transitions from the
continuum to levels with {\it n} = 6 to 9.  The resulting intrinsic flux 
densities expected from the entire nebula, combining these thermal processes,
are 4, 8, 1, and 2~mJy in IRAC Bands~1,2,3,4, respectively. We now require
an estimate of the intervening extinction to assess whether these processes
make any observable MIR contributions to the nebular emission.

The thermal radio emission provides an estimate of the intrinsic emission
line flux in H$\alpha$ (see Condon (1992), eqns. (3) and (4a)).  Using this
formulation yields 2.1$\times10^{-11}$~erg~cm$^{-2}$~s$^{-1}$ for the integrated 
intrinsic H$\alpha$ flux, before the effects of extinction. To determine an upper limit
for the spatially-integrated H$\alpha$ flux of the nebula, we have applied the
absolute calibration procedure to the SHS H$\alpha$ image that Pierce (2004)
describes.  The SHS obtains short-red (SR) continuum exposures of each H$\alpha$
field.  Numerically scaled versions of these were intended to be used to remove 
any continuum that is included in the narrowband H$\alpha$ filter.
After detailed cross-comparison with the absolutely calibrated, but spatially much 
lower-resolution, Southern H$\alpha$ Sky Survey Atlas (Gaustad et al.
2001), Pierce provides both the SR scale factors and the number of flat-fielded 
SHS H$\alpha$ counts pixel$^{-1}$ that are equivalent to 1~Rayleigh 
($10^{-7}$~erg~cm$^{-2}$~s$^{-1}$~sr$^{-1}$ at H$\alpha$).  The ring lies in the 
SHS field HAL0182, for which Pierce gives an SR scale factor of 1.12 and 
7.8~counts pixel$^{-1}~R^{-1}$, although there is a very large amount of
scatter about the latter value in this field (Pierce 2004, Table~2).  Visual 
examination of this calibrated SHS H$\alpha$ image, guided by overlaying the 13-cm 
outer radio contour of the nebula,
suggests that a potential excess of H$\alpha$ emission exists within the ring.  
However, there is no statistically significant excess compared with other regions of
the image lacking point sources.  Taking the measured H$\alpha$ flux within
the ring as a conservative upper limit to the H$\alpha$ emission, the
brightness integrated over the nebula is less than 9.0$\times10^{-14}$~erg~cm$^{-2}$~s$^{-1}$.
Comparing this limit with the predicted H$\alpha$ emission implies that A$_V$ exceeds 13$^m$
(e.g. using the reddening law of Cardelli et al. (1989) in the optical,
assuming A$_{V}$/$E(B-V)$~=~3.1 (appropriate for the diffuse interstellar medium)).
Using the extinction maps of Schlegel et al. (1998), the total Galactic
extinction along the line-of-sight to the nebula is A$_V$$\sim$36$^m$.  Although this
is large it is not atypical for the cores of spiral arms.

The influence on IRAC flux densities of this substantial visual extinction cannot be neglected.
Adopting an A$_V$ of $36^m$ yields extinctions ($\pm1\sigma$) in the IRAC bands of 2.28$\pm$0.25, 
1.75$\pm$0.32, 1.75$\pm$0.40, and 1.75$\pm$0.40$^m$, respectively, from the empirical reddening 
law found by Indebetouw et al. (2005) for the IRAC bands along two lines-of-sight.  A conventional 
reddening law (e.g. Cohen et al. 2003) gives extinctions of 1.85, 1.26, 0.94, and 1.59$^m$, 
respectively.  Both these sets of attenuations would render the expected contributions of gaseous
thermal processes to the observed flux densities negligible (1~mJy or less) in all IRAC bands.

The radio flux density can be converted into the ionized mass of the ring (Mezger \& 
Henderson 1967, Appendix A, eqn. A.14), using T$_e$~=~10$^4$~K, assuming the
simplest model of a uniformly filled sphere, and adopting a 5-GHz flux density of
25~mJy, equal to the best fit to the three lower frequencies described above.  The ionized
mass is 0.0022$\times$D$_{kpc}^{2.5}$~M$_\odot$, where D$_{kpc}$ is the distance to 
the ring in kpc.  The mean ionized mass of planetary nebulae (PNe) 
is between 0.1 and 0.25~M$_{\odot}$ (Boffi \& Stanghellini 1994), 
suggesting that the ring is between 4.6 and 6.6~kpc from the Sun.  
In this direction the line-of-sight extinction
based on reddened OBA stars amounts to only $\sim$3$^m$ out to a distance of 3~kpc 
(Neckel et al. 1980). The nebula is seen in the direction of the tangent 
to the Scutum-Crux spiral arm.  The tangent point lies 6.6~kpc 
from the Sun (Taylor \& Cordes 1993).   We, therefore, favor a distance close to 6.6~kpc
because it provides a natural explanation for the extremely high extinction that the 
PN suffers at a relatively modest distance from the Sun, 
namely in the dust lane associated with 
this arm.  Identifying the reddening with an arm links the bulk of the DIRBE-derived 
extinction in this longitude to the spiral arm, rather than being distributed along 
an extended line-of-sight through the Galaxy as, for example, the 30$^m$ of A$_V$ 
associated with the direction to the Galactic center.  The implied nebular radius
is 0.4~pc, close to the median (0.3~pc) for a set of PNe in the Magellanic Clouds
(Stanghellini et al. 1999).

\section{Discussion}
The morphology of the MIR and radio continuum images suggests that ionized gas
dominates the 4.5-$\mu$m image, while PAHs contribute dominantly at 8.0-$\mu$m.
However, as demonstrated above, the contributions to the 4.5-$\mu$m emission 
from thermal processes that involve atomic hydrogen are very small.  Therefore, any
ionic lines within this band must come from heavy elements with rather
strong {\rm gf} values; for example, [Mg{\sc iv}] at 4.485~$\mu$m, and [A{\sc vi}] 
at 4.530~$\mu$m.  Both these lines are seen in high-excitation PNe.  In NGC~7027
and NGC~7662, which have significant MIR continua, the lines account for 
almost 10\% of the flux in the 4.5-$\mu$m band. In a PN lacking such a continuum 
the lines' contribution could be larger.  Alternatively, a variety of H$_2$
emission lines could contribute in several IRAC bands, but chiefly at 8.0~$\mu$m, 
as Hora et al. (2004b) found from an IRAC survey of nine PNe.  

Were the nebula to be a supernova remnant (SNR) we would expect a negative radio
spectral index.  Even if the SNR were young, such a shell source would have a
negative index reflecting the synchrotron emission mechanism.  Examples
of young SNRs with a ringlike structure include Tycho, Kepler, Cas~A, and SN~1006,
and all their radio indices are around $-0.7$. Accretion onto a compact object can
produce a flat spectrum, as could a plerion.  
The $13-20$~cm spectral index ($\sim$1.7) for the central 
source is rather uncertain, but probably excludes a flat spectrum, making this 
interpretation unlikely.  If the central source were a very young 
pulsar wind nebula, the spectrum could even be inverted as in SN 1986J, 
which recently showed a double power law with slopes $-0.55$ and $+1.38$ (Bietenholz 
et al. 2002).  However, this time-dependent inversion appears only
at high radio frequences ($>$10~GHz).  These authors found synchrotron 
self-absorption within the source was an unlikely explanation for the steeply 
positive power law.  The ring's radio emission is too weak to measure polarization 
reliably with current instrumentation.  

\begin{figure}
\plotone{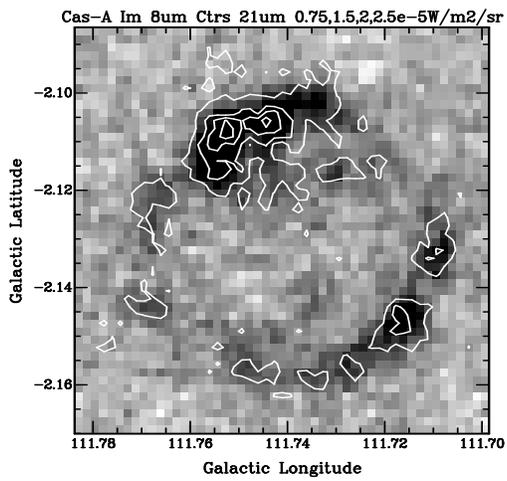} 
\figcaption{MSX 8.3-$\mu$m greyscale image of Cas~A overlaid by white contours of
21.3-$\mu$m emission at values of 0.75, 1.5, 2, 2.5$\times$10$^{-5}$~W~m$^{-2}$~sr$^{-1}$.
The lowest contour is 3.5$\sigma$ above the background.\label{casa}}
\end{figure}

The MIR morphology is also very different from that of a
young SNR, e.g. Cas~A.  Figure~\ref{casa} illustrates the MSX data for this  
young remnant, overlaying 21.3-$\mu$m contours on the 8.3-$\mu$m image.  The
SNR is highly clumpy, shows large apparent density contrast and ragged inner and  
outer edges.  These characteristics are unlike those of G313.3+00.3.  
Recent imaging with the {\it Spitzer} Space Telescope (Hines et al.
2004) demonstrates that MIR emission in Cas~A correlates well with X-ray and
optical emission lines rather than with synchrotron regions.  These authors
attribute the MIR energy to thermal emission from dust inside the SNR, or associated 
with emission-line gas inside the reverse shock region.  This correlation may 
arise from grain formation in regions with sufficiently cool gas, militating against
a smoothly varying distribution of MIR emission.

Nonthermal emission in the IR and radio would lead to a ratio of 
$F_{8.3 \mu m}/S_{843 MHz}$~$\leq$~3 (Cohen \& Green 2001).  In fact, the nebular 
ratio is about 6 for G313.3+00.3.  This is closer to the values found for PNe, 
for which the median ratio is 12, with a semi-interquartile range of 6, than 
to H{\sc ii} regions (median 24). If the MIR spectrum were that of a PN or an SNR 
dominated entirely by emission lines, as in high-excitation objects, one would 
expect spatial variations across the nebula in the intensity ratios among the IRAC 
bands (implicit in Figure~\ref{casa}).  We now investigate the integrated intensity ratios of 
8.0-$\mu$m/5.8-$\mu$m and 8.0-$\mu$m/3.6-$\mu$m in G313.3+00.3 and compare these with known PNe.
If it were a PN in which PAHs dominated the spectral energy distribution (SED) then 
band-to-band ratios should be rather uniform across the nebula.  We eliminated the 
contributions from the presumed unrelated, bright point source within the ring, and 
from the central star.  The nebular mean intrinsic values for these two ratios in
G313.3+00.3, derived by dereddening the observed maps, are $3.1\pm0.2$ and $12\pm3$, and
are quite uniform over the nebula.  These uncertainties incorporate the differences in 
corrections between a conventional extinction law and that presented by Indebetouw et al. 
(2005).

A rough estimate of what to expect for a PN with only PAH emission 
comes from assigning the observed 3.6, 5.8, and 8.0-$\mu$m PN 
intensities purely to the 3.3, 6.2, 7.7, and 8.7-$\mu$m PAH bands.  The predicted ratios 
would then be 2.7$^{+0.5}_{-0.3}$ (Band~4/Band~3) and 17$^{+4}_{-3}$ (Band~4/Band~1) (Cohen et al. 
1986: Table 2, for PNe; Cohen et al. 1989: Tables 3 and 4, for PNe).  
These are very similar to the ratios for G313.3+00.3.  A more precise
approach is to synthesize IRAC photometry from MIR spectra of PNe.  Volk \& Cohen (1990)
characterized the 7.7 to 22.7-$\mu$m spectra of 170 PNe according to the dominance of PAHs,
broad or narrow dust emission features, and of atomic fine structure lines of low, high,
or very high excitation.  The recognition that, in some bands, almost all the MIR emission in a 
PN can be contributed by pure rotational lines of H$_2$ (e.g. Cox et al. 1998) implies 
that one must additionally consider H$_2$ line emission.  To quantify this emission in an extreme case
of a PN entirely dominated by H$_2$ lines, we have combined the line data for the Helix Nebula (Cox et 
al. 1998) with those for NGC~6302, NGC~7027, and Hb~5 (Table~7 of Bernard-Salas \& Tielens (2005)).  In 
real PNe, the H$_2$ line emission that falls within the IRAC bands is also accompanied by MIR 
continuum, fine structure lines, and/or PAHs.  We have integrated the new (July 2004) IRAC relative 
spectral response curves over the spectra of 63 PNe observed with the ISO Short Wavelength Spectrometer
(including three of the H$_2$-emitting PNe cited above), and have
formed the observed intensity ratios 8.0-$\mu$m/5.8-$\mu$m and 8.0-$\mu$m/3.6-$\mu$m.
Allowance for extinction in these PNe would change the observed ratios very little.  (For a PN
with a modest A$_V$ of 3$^m$, the intrinsic ratios would differ from those observed by at most 4\% 
percent, whichever reddening law is used.  In fact,  Band~4/Band~3 is unaltered with the 
Indebetouw et al. (2005) extinction.) We have averaged the ratios observed for PNe grouped
according to their most conspicuous MIR spectral character.  

For 16 PNe dominated by PAH emission the mean ratios of 8.0-$\mu$m/5.8-$\mu$m and 
8.0-$\mu$m/3.6-$\mu$m are found to be 2.9$\pm$0.2 and 14$\pm$3, entirely consistent (within
1$\sigma$) with
those of G313.3+00.3.  Could another emission mechanism mimic these radiance ratios?  The
most restrictive criterion is the 8.0-$\mu$m/3.6-$\mu$m ratio.  Only for PNe dominated by
very high excitation emission lines does this ratio attain a value (14) comparable to that of
G313.3+00.3 but, for such PNe, 8.0-$\mu$m/5.8-$\mu$m is much too low ($\sim1$).  No single
emission process other than PAHs can simultaneously match the intrinsic Band~4/Band~3 and 
Band~4/Band~1 ratios of G313.3+00.3.  Specifically, H$_2$ line emission alone 
gives mean ratios of 5.5$\pm$1.2 and 4.9$\pm$1.8, respectively, quite inconsistent
with G313.3+00.3.  Might a combination of emission processes match the ratios in G313.3+00.3? 
Adding the pure H$_2$ to the very high excitation line spectrum in the proportion 
of 1:3 would formally produce spatially integrated nebular ratios of 2.8$\pm$0.5 and 
11.9$\pm$2.8.  However, it is impossible for such values to be relatively uniform
across a PN because each process 
requires radically different physical conditions.  H$_2$ emission arises near the surface
of PDRs but the chief cooling lines are in the far-IR, from low excitation ions 
such as [C{\sc ii}], [O{\sc i}], and  [Si{\sc ii}] (Bernard-Salas \& Tielens 2005), not the 
species that characterize IRAC's wavelength range in very high excitation PNe such as 
NGC~2440 (Bernard-Salas et al. 2002), e.g. [Ar{\sc vi}] and [Ne{\sc vi}].  The band-to-band 
ratios of G313.3+00.3 are, therefore, consistent only with a PAH-dominated spectrum.

\begin{figure}
\plotone{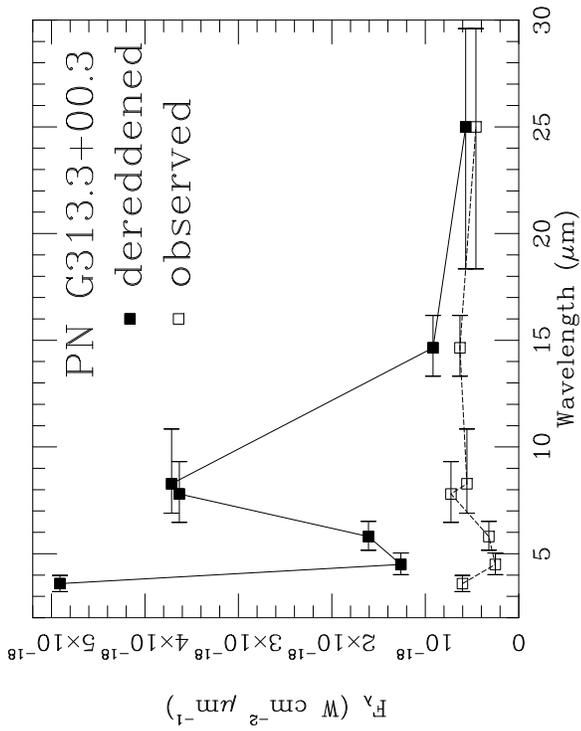} 
\figcaption{The observed (open squares, dashed line) and intrinsic (filled squares,
solid line) SEDs for the PN before, and after, correction for 36$^m$ of visual extinction.  
Horizontal bars through each data point indicate the FWHM of the associated filter bands.
Note the clear indications of PAH emission at 3.6 and 8.0~$\mu$m.\label{sed}}
\end{figure}

Figure~\ref{sed} shows the spatially-integrated SED for the nebula, both observed, and after 
dereddening corresponding to an A$_V$ of $36^m$, using Indebetouw et al. (2005).  For each data point,
asymmetric horizontal bars represent the extent of the FWHM of the corresponding filter band. The 
obvious features appear to correspond with PAH emission bands at 3.3 and 7.7~$\mu$m.  Only a weak 
PAH contribution would be expected in the 14.6-$\mu$m MSX band from C-C-C modes near 16~$\mu$m.  
The dominance of the 3.3-$\mu$m PAH band in this plot of specific intensity is misleading.  
Both this feature and the IRAC 3.6-$\mu$m bandwidth are far narrower than the 7.7~$\mu$m band
and the 8.0-$\mu$m filter, greatly reducing the integrated in-band flux contributed by PAHs 
to the IRAC Band-1 data point as compared with Band-4.  Between 3 and 30~$\mu$m the ring emits 
only about 700~L$_{\odot}$, even after dereddening by 36$^m$.  However, the complete absence of 
any optical or NIR detection of the ring renders any meaningful estimate of its bolometric luminosity 
impossible from our observations.  The bulk of the expected luminosity should occur in the optical 
and UV, but we lack any data to deredden at these short wavelengths.  

Scaling down the predicted IR thermal emission of the entire nebula to that of 
the central source by the ratio of their 13-cm flux densities suggests approximately
0.2 and 0.4~mJy in IRAC Bands~1 and 2.   These predictions are of the same order as 
the values measured for the central star, and the predicted flux density ratio is in 
accord with that observed, at the 1$\sigma$ level.  
(In the above simplistic calculation we have neglected the probable higher values of
T$_e$ and ionic charge expected within the stellar wind as opposed to the overall
nebula.) 
It is somewhat unusual for a star 
to have a radio counterpart.  But this a plausible explanation in the 
context of the central star of a PN.  Taylor et al. (1987) modeled
thermal radio emission from such a stellar wind with a 1/r$^2$ density profile,
and matched the observed radio spectra for a sample of PNe with slopes of $\ge$1.
Truncation of these stellar winds also steepens radio spectral indices (Marsh 1975).
Therefore, the central star, or at least its wind, is detected in both 
the IR and the radio.

We conclude that G313.3+00.3 is most likely to be a PN within heavy
obscuration associated with our almost tangential view along the Scutum-Crux
spiral arm.  The regular nebular morphology, the presence of a radio and IR source 
at the geometric center, the spatially rather uniform ratio-images in key IRAC bands, 
and the quantitative match to what is expected for these band ratios from PAH emission, 
all support this classification.  We anticipate that this PN will be only the first 
of many highly extinguished planetary nebulae to be discovered by GLIMPSE.  

\acknowledgments
The Australia Telescope is funded by the commonwealth of Australia for operation
as a National Facility managed by the CSIRO.  The MOST is owned and operated by
the University of Sydney, with support from the Australian Research Council and
Science Foundation within the School of Physics.
Support for this work, part of the {\it Spitzer} Space Telescope Legacy Science 
Program, was provided by NASA through contracts 1224653 (Univ. Wisconsin Madison), 
1225025 (Boston Univ.), 
1224681 (Univ. Maryland), 1224988 (Space Science Institute), 1259516 (UC Berkeley). 
MC also thanks NASA for support under ADP grant NNG04GD43G with UC Berkeley.  
This research made use of Montage, funded by the
National Aeronautics and Space Administration's Earth Science Technology
Office, Computational Technnologies Project, under Cooperative Agreement
Number NCC5-626 between NASA and the California Institute of Technology.
This research made use of data products from the Midcourse Space
eXperiment.  Processing of the data was funded by the Ballistic
Missile Defense Organization with additional support from the NASA
Office of Space Science.  This research has also made use of the
NASA/IPAC Infrared Science Archive, which is operated by the
Jet Propulsion Laboratory, California Institute of Technology,
under contract with the National Aeronautics and Space Administration.

\clearpage

\clearpage

\clearpage

\clearpage

\clearpage

\clearpage

\clearpage

\clearpage

\clearpage
\begin{deluxetable}{ll} 
\tablecolumns{2} 
\tablewidth{0pc} 
\tablecaption{Radio and IR flux densities of the ring and central star.\label{flux}} 
\tablehead{Waveband &  Flux Density\\}
\startdata
{\bf The nebula} & \\
MOST 843~MHz [mJy]            & $ 23\pm3  $ \\
ATCA 1384~MHz (low-resolution) [mJy]            & $ 35\pm5  $ \\
ATCA 2496~MHz (low-resolution) [mJy]            & $ 25\pm3    $ \\
MSX/CB-02 F(8.3$\mu$m) [mJy]            & $ 130 $  \\
MSX/CB-03 F(8.3$\mu$m) [mJy]            & $ 125 $  \\
MSX/CB-03 F(14.6$\mu$m) [mJy]            & $ 450 $  \\
{\it IRAS} F(12$\mu$m) (Jy)           & $<2.2 $    \\
{\it IRAS} F(25$\mu$m) (Jy)           & $0.81\pm0.15$    \\
{\it IRAS} F(60$\mu$m) (Jy)            & $<8.0 $    \\   
{\it IRAS} F(100$\mu$m)(Jy)            & $<140 $    \\
IRAC F(3.6$\mu$m [mJy]            & $ 30 $  \\
IRAC F(4.5$\mu$m [mJy]            & $ 25 $  \\
IRAC F(5.8$\mu$m [mJy]            & $ 38 $  \\
IRAC F(8.0$\mu$m [mJy]            & $ 150 $  \\
{\bf The Central Star} & \\
ATCA 1384~MHz (high-resolution) [mJy]            & $ 2.6:  $ \\
ATCA 2496~MHz (high-resolution) [mJy]            & $ 1.0\pm0.05    $ \\
IRAC F(3.6$\mu$m [mJy]            & $ 0.30\pm0.06 $  \\ 
IRAC F(4.5$\mu$m [mJy]            & $ 0.43\pm0.08 $  \\ 
\enddata 
\end{deluxetable} 

\end{document}